\documentclass[journal]{IEEEtran}
\usepackage{amsmath,amsfonts}
\usepackage{algorithmic}
\usepackage{array}
\usepackage[caption=false,font=normalsize,labelfont=sf,textfont=sf]{subfig}
\usepackage{textcomp}
\usepackage{stfloats}
\usepackage{url}
\usepackage{verbatim}
\usepackage{graphicx}
\def\BibTeX{{\rm B\kern-.05em{\sc i\kern-.025em b}\kern-.08em
    T\kern-.1667em\lower.7ex\hbox{E}\kern-.125emX}}
\usepackage{balance}

\begin{document}

\title{Dissociable Spatial and Temporal Effects of Interaction Latency in Virtual Reality}

\author{
    Xiaoye Michael Wang, %
    \thanks{X. M. Wang is with the University of Toronto, Toronto, ON, Canada, and can be contacted at michaelwxy.wang@utoronto.ca.}%
    Catherine M. Sabiston, %
    \thanks{C. M. Sabiston is with the University of Toronto, Toronto, ON, Canada, and can be contacted at catherine.sabiston@utoronto.ca.}%
    Timothy N. Welsh %
    \thanks{T. N. Welsh is with the University of Toronto, Toronto, ON, Canada, and can be contacted at t.welsh@utoronto.ca.}%
}

\markboth{Wang \MakeLowercase{\textit{et al.}}: Dissociable Spatial and Temporal Effects of Interaction Latency in Virtual Reality}%
{Wang \MakeLowercase{\textit{et al.}}: Dissociable Spatial and Temporal Effects of Interaction Latency in Virtual Reality}

\maketitle

\begin{abstract}
    Motion-to-photon latency is inherent in immersive virtual reality (VR) systems and can arise from multiple sensorimotor loops, including view-contingent latency between head movement and display update and interaction latency between hand movement and the displayed virtual effector. Although previous work has shown that interaction latency can impair VR performance, it remains unclear whether commonly used spatial, temporal, and efficiency measures reveal the same latency-related disruption. The present study addressed this question by experimentally imposing delays between the physical and virtual hands during manual pointing in immersive VR. Participants pointed to targets on a horizontal surface in VR and in the physical environment as an unmediated baseline. In VR, pointing was performed with a virtual hand avatar controlled by an external motion capture pipeline, and additional delays (0--500 ms) were imposed between the participant's physical hand movement and the rendered movement of the virtual hand. Relative to the unmediated physical baseline, performance in VR showed greater endpoint error, longer movement time, greater endpoint variability, and lower throughput. Within VR, added interaction latency further increased endpoint error and variability, reduced throughput, and altered movement time, but these effects followed different profiles: endpoint error increased even at the shortest delays, whereas movement time remained relatively stable at short delays and increased primarily at longer delays. These findings show that interaction latency produces dissociable spatial and temporal consequences in immersive VR, such that endpoint accuracy revealed disruption before movement time or throughput. Thus, latency-sensitive VR interactions cannot be fully evaluated using movement time or efficiency measures alone. Instead, HCI evaluations should assess both spatial and temporal performance, particularly when VR tasks involve visually guided manual actions.
\end{abstract}


\section{Introduction}

Between the late 2010s and 2020s, rapid advancements in computer hardware and machine learning technologies have expedited the development of extended reality (XR) technologies, including virtual reality (VR) and augmented reality (AR). This general trend has sparked broad interest in exploring the application scenarios afforded by XR technologies beyond gaming, including rehabilitation (e.g., \cite{Elor2020, Naqvi2024}), medical (e.g., \cite{Giraldo2025}) and laboratory (e.g., \cite{Wang2025ASEE}) training, and coaching in sports (e.g., \cite{Yin2025}). As a result, XR technologies and their applications have strong potential to transform digital interaction beyond the classic WIMP (Windows, Icons, Menus, Pointer) desktop paradigm by moving user interaction from traditional two-dimensional (2D) interfaces into immersive three-dimensional (3D) spaces \cite{WangInPressHCIHandbook}.

XR technologies also come with unique shortcomings and challenges that warrant careful evaluation before widespread adoption because their unique hardware characteristics may inadvertently produce unintended negative effects. For instance, head-mounted display (HMD)-based VR is commonly treated as a proxy for physical, unmediated reality (UR) because it simulates the visual experience of UR via digital displays \cite{Wang2023VisualCognition,Wang2024JOV}. Although the digital renderings recreate multiple forms of optical information needed for perception and action, the displays' fixed focal distance typically results in the vergence-accommodation conflict (VAC), which yields depth compression \cite{Hussain2023,Wang2026Displays,Wang2024bVR}, degrades 3D selection performance \cite{Batmaz2022,Batmaz2023}, affects users' perception of the size of their virtual avatar \cite{Wang2026FunctionalBodySize}, and, crucially, temporarily transfers to and degrades performance in the physical environment \cite{Wang2024aSciRep}. These findings suggest that VR should not be assumed to function as a neutral proxy for UR, nor should perceptuomotor skills acquired in VR be expected to transfer veridically to UR without careful empirical validation \cite{ManzoneUnderReview, WangInPressHCIHandbook, Wang2025BBR}.

Another hardware-related constraint that can influence perception and action in XR is latency, the temporal lag between a user's movement and the corresponding update of the virtual environment. Latency is a widely studied topic in human-computer interaction (HCI) research \cite{Attig2017}. Due to the finite processing speed of computing devices, this latency may never reach zero \cite{Papadakis2011}. In 2D interfaces, increased latency has been shown to affect both user experience \cite{Fischer2005,Zhou2016} and performance \cite{Butler1983,Friston2015,Martens2018}. In VR, users' head and sometimes limb movements are tracked using either external or built-in sensors, after which the tracking data are transferred and processed to render the virtual environment as stereoscopic images on the HMD and to animate the virtual avatar for interaction. The cumulative delay across this processing and rendering pipeline is termed motion-to-photon latency \cite{Warburton2023}. Within the broader category of latency, the present study focuses specifically on interaction latency -- the delay between a user's hand movement and the corresponding update of the virtual effector. Both motion-to-photon latency in general and interaction latency in particular have been shown to affect VR task performance \cite{Hoyet2019,Waltemate2016} and, in some cases, contribute to cybersickness \cite{Lackner2000,Stauffert2018,Stauffert2020}.

Warburton et al. \cite{Warburton2023} reported that different VR systems exhibit different latency profiles. From an HCI perspective, such variability makes it difficult to determine whether differences in user performance (e.g., movement time or accuracy) should be attributed to task demands or to device-specific delays. Moreover, for VR applications that require rapid and spatially accurate interactions, such as surgical simulation \cite{Giraldo2025}, laboratory or engineering training \cite{Wang2025ASEE}, sports training (e.g., hockey training \cite{Buns2020}), and teleoperation (e.g., drone pilot training \cite{CardonaReyes2021}), even modest fluctuations in latency can directly influence the effectiveness of these platforms. Therefore, as with VAC, it is important to systematically characterize how latency affects user performance.

Although there are established methods to measure motion-to-photon latency in VR systems (e.g., \cite{Warburton2023,Zhao2017}), it remains difficult to determine how increasing interaction latency alters manual performance in immersive VR, particularly when performance is compared with the physical environment and when spatial and temporal consequences are examined separately. This problem arises because any VR-physical difference reflects not only added interaction latency, but also other changes inherent to immersive VR, including altered visual rendering (e.g., VAC; \cite{Wang2024bVR}), transformed effector representation (e.g., \cite{ManzoneUnderReview, venkatakrishnan2023virtual}), and disrupted sensorimotor coupling (e.g., \cite{Wang2025BBR}). Consequently, it is difficult to isolate the specific contribution of latency to observed changes in endpoint accuracy and movement timing, which may themselves be affected through partly distinct mechanisms. This challenge is further compounded by the fact that native system latency varies across devices and across the time course of a movement \cite{Warburton2023}. Although it might not be possible to eliminate these latencies, one way to begin addressing this problem is to gain insight into the perceptuomotor impacts of different delays  systematically manipulate interaction latency within a controlled VR setup and examine the resulting changes in performance relative to a baseline VR condition with native system latency and to UR \cite{Waltemate2016}.

To isolate the specific contribution of interaction latency to manual performance in immersive VR, the current study systematically examined how imposed rendering delays shape targeted manual pointing movements. Using a controlled virtual hand setup in which additional rendering delays were imposed on top of a stable baseline pipeline, the study isolated the effects of added interaction latency by comparing performance both across VR latency conditions and against performance in UR. In doing so, the study quantified how latency influences movement accuracy and timing during targeted manual actions. The present study makes two main contributions:
\begin{itemize}
\item It shows that interaction latency produces dissociable spatial and temporal consequences in immersive VR, such that endpoint accuracy reveals disruption before movement time or throughput do.
\item It provides guidance for HCI evaluation by showing that latency-sensitive VR interactions should be assessed using both spatial and temporal performance measures rather than either measure alone.
\end{itemize}
Together, these findings provide a basis for understanding how interaction latency shapes perceptuomotor behavior and for informing the design and evaluation of VR systems.

\section{Related Work}

\subsection{Latency and User Interaction}

Latency has long been recognized as a central issue in HCI, not only because latency shapes users' subjective impressions of responsiveness, but also because it can directly affect task performance \cite{Caldwell2009,MacKenzie1993}. Early work already showed that response time of the interface can alter user performance, establishing latency as a functional constraint rather than merely a source of annoyance \cite{Butler1983}. For this reason, both classic and more recent guidelines have emphasized the importance of keeping delays between user action and system response as short as possible \cite{Attig2017}. At the same time, because latency is introduced by sensing, computation, transmission, and display processes, truly zero delay is rarely achievable in practice. As a result, HCI research has examined how latency effects vary with delay magnitude, user expectations, task complexity, and interaction context, rather than treating latency as having a single universal threshold \cite{Attig2017,Dabrowski2011,Doherty2015,Shneiderman1984}. Empirical work further shows that delay influences perceived responsiveness and usability \cite{Fischer2005}, and that tolerance for delay varies with interface characteristics and task demands \cite{Zhou2016}. Together, this literature indicates that latency does not affect interaction uniformly: its consequences depend on the task and on how tightly user action must be coupled to system feedback.

Prior HCI work has also shown that latency produces specific behavioral consequences for action performance. In pointing and steering tasks, Friston et al. \cite{Friston2015} reported that movement time costs emerged once delay exceeded a functionally meaningful range, with little effect at very short delays but clear slowing at larger values. Even relatively small delays can be behaviorally meaningful. Martens et al. \cite{Martens2018} showed that low-range latency (14--198 ms) affected both performance and perception in a virtual ball-balancing task that required continuous stabilization of an unstable system. These studies suggest that latency can impair user interaction by altering responsiveness, changing movement timing, and degrading task performance once delay becomes sufficiently large relative to the temporal demands of the task.

\subsection{Motion-to-Photon and Interaction Latency in VR}

In VR systems, latency can directly affect user experience and interaction quality \cite{Caserman2019}. In immersive VR, users expect their actions to produce timely and spatially consistent changes in visual feedback. This action-perception coupling is a form of sensorimotor contingency \cite{ORegan2001}, and can contribute to presence, or the sense of being in the virtual environment \cite{Slater2009,Slater2022}. When this coupling is perturbed by system latency, the virtual environment may no longer respond as a stable extension of the user's movements, with consequences for perception, action, and comfort.

Similar to latency in conventional interfaces, latency in VR is not a unitary phenomenon. For the current study, it is useful to distinguish between two functional sources of latency in VR: view-contingent latency and interaction latency. View-contingent latency refers to the delay between head movement and the corresponding update of the rendered viewpoint \cite{LaValle2014}. This form of latency is closely tied to perceptual stability because the visual world must remain properly coupled to head motion. Allison et al. \cite{Allison2001} showed that end-to-end display lag in head-coupled virtual environments degrades the perceived stability of the visual world, especially as head velocity increases. Delay in head-coupled displays has also been linked to simulator sickness and related perceptual disturbances \cite{Draper2001,Lackner2000,Stauffert2020}. For this reason, view-contingent latency is typically treated as a display-level constraint that should be minimized.

A second source is interaction latency, that is, delay between the user's hand or controller movement and the corresponding update of the virtual effector or interaction outcome. This form of latency is especially relevant for manual action because it alters the temporal relationship between movement and its visual consequences. In an early reaching study using a head-coupled stereo display, Ware and Balakrishnan \cite{Ware1994} reported that hand-tracking lags as low as 87 ms significantly impaired target acquisition performance, whereas head-tracking lag had little effect, suggesting that interaction latency is especially consequential for visually guided object manipulation. Commercial VR systems often use motion prediction to reduce effective latency by rendering based on predicted future motion rather than relying only on delayed tracking samples \cite{Hou2020,Hou2019,LaValle2014}. However, because such prediction depends on already observed movement, the benefit of prediction can vary over the course of an action, particularly near movement onset or during rapid changes in trajectory. Consistent with this issue, Warburton et al. \cite{Warburton2023} emphasized the need to measure motion-to-photon latency specifically for controller-based sensorimotor experiments, because effective latency can vary across systems and across phases of a movement. Thus, interaction latency is not always a single fixed quantity in practice, further motivating controlled experimental manipulations when relating latency to manual performance in VR.

\subsection{Effects of Interaction Latency in VR}

For manual interaction in VR, the key question is how delayed visual feedback alters visually guided movement. Prior work suggests that interaction latency does not affect all perceptual and behavioral outcomes equally. Waltemate et al. \cite{Waltemate2016} found that motor performance and simultaneity perception deteriorated at delays above 75 ms, whereas agency and ownership remained relatively robust until delays exceeded 125 ms, suggesting that action-related measures are more latency-sensitive than embodiment-related judgments. Hoyet et al. \cite{Hoyet2019} similarly found that faster avatar movements lowered latency detection thresholds and increased performance costs. More recently, Yang et al. \cite{Yang2025} showed that latency perception thresholds in a 3D VR pointing task varied with interaction strategy and task difficulty, with faster or more demanding interactions producing greater sensitivity to latency. Together, these findings indicate that the effects of interaction latency depend on movement demands and on the particular outcome being measured.

This task- and outcome-dependence creates a measurement challenge. The effects of latency in pointing and selection tasks are often summarized using movement time, throughput, or selection-level accuracy measures such as error rate or success rate. These metrics are useful for characterizing overall interaction efficiency, but they may obscure how latency alters the spatial structure of the movement itself, such as signed endpoint bias and endpoint variability. This distinction is especially relevant as VR increasingly supports applications that require continuous spatially guided actions rather than discrete target selection alone. In many traditional selection tasks, performance is ultimately evaluated by whether a target is successfully acquired \cite{Amini2025,ISO2015}, making the task relatively tolerant to substantial spatial deviations within the target boundary. By contrast, applications such as surgical simulation often depend on the spatial fidelity of the movement itself. In these contexts, latency-induced changes in endpoint bias may be behaviorally meaningful even when overall task completion or selection success remains unaffected.

One reason for this discrepancy is that latency may alter the spatial and temporal structure of a movement without necessarily changing the final task outcome. For example, increased movement time may reflect either a direct temporal consequence of delayed visual feedback or strategic slowing to preserve accuracy through additional late-stage correction near the target. Yang et al. \cite{Yang2025} decomposed VR 3D pointing into reaction, ballistic, and correction phases. They found that latency had little effect on reaction time, with ballistic phase duration increasing approximately linearly with latency and correction phase duration increasing nonlinearly at higher delays. This phase-specific pattern suggests that latency may affect early movement execution and late corrective control differently, rather than simply slowing the movement as a whole. 

This issue is especially important for targeted manual pointing in immersive VR because accurate performance depends on both spatial and temporal control. Prior 3D VR pointing work has shown that Fitts' law can be extended to immersive environments, but additional factors such as depth perception, movement direction, display characteristics, and virtual-hand feedback introduce variability that can reduce the predictive accuracy of models derived from 2D pointing tasks \cite{Amini2025,BarreraMachuca2019,Batmaz2019,Clark2020,Janzen2016}. Consequently, measures that primarily emphasize task completion, movement time, or overall efficiency may not fully capture how latency alters the spatial fidelity of visually guided movements.

The present study builds on this work by experimentally manipulating interaction latency during targeted manual pointing in immersive VR and comparing performance against both a baseline VR condition and an unmediated physical baseline. The key question was not simply whether latency degrades performance, but whether different performance measures reveal the same latency effect. By jointly examining endpoint accuracy and variability, movement time, and throughput, this study tested whether interaction latency produces dissociable spatial and temporal consequences within the same task.

\section{Research Questions and Hypotheses}

To clarify how interaction latency affects manual performance in immersive VR, the current study examined targeted pointing movements under experimentally manipulated delays between the participant's hand movement and the displayed virtual hand. The study was guided by the following research questions:

\begin{description}
    \item[\textbf{RQ1}:] How does increasing interaction latency alter targeted pointing performance in VR relative to the physical environment?
    \item[\textbf{RQ2}:] Do different performance measures reveal the same latency-related disruption, or do spatial and temporal measures respond differently to increasing interaction latency?
\end{description}

Based on prior work showing that delayed visual feedback can degrade visually guided action, and on the premise that interaction latency disrupts the coupling between movement and its visual consequences, three \textit{a priori} hypotheses were tested:
\begin{description}
    \item[\textbf{H1}:] Relative to performance in the physical environment, targeted manual pointing in native VR will show degraded performance, including longer movement times and reduced endpoint accuracy, even when no additional interaction latency is introduced.
    \item[\textbf{H2}:] Within VR, increasing additional interaction latency will further impair targeted manual pointing performance beyond the baseline performance cost observed in native VR.
    \item[\textbf{H3}:] Spatial and temporal performance measures will show dissociable response patterns as interaction latency increases.
\end{description}

\section{User Study}

\subsection{Participants}

Twenty-two (22) right-handed adults (16 males, 6 females) between 18 and 23 years of age (mean = 20.77 years, SD = 1.23) participated in the study. All participants reported normal or corrected-to-normal vision with no known neurological or motor impairments. Participants provided full, written informed consent prior to their participation in the experiment. All procedures were approved by the University of Toronto Ethics Review Board. All participants were naïve to the experimental hypotheses and had minimal prior VR experience. Two participants' data were discarded due to technical challenges during data collection, leaving a final sample of 20 participants. 

\subsection{Stimuli and Apparatus}

\begin{figure}[tb]
  \centering
  \includegraphics[width=\columnwidth]{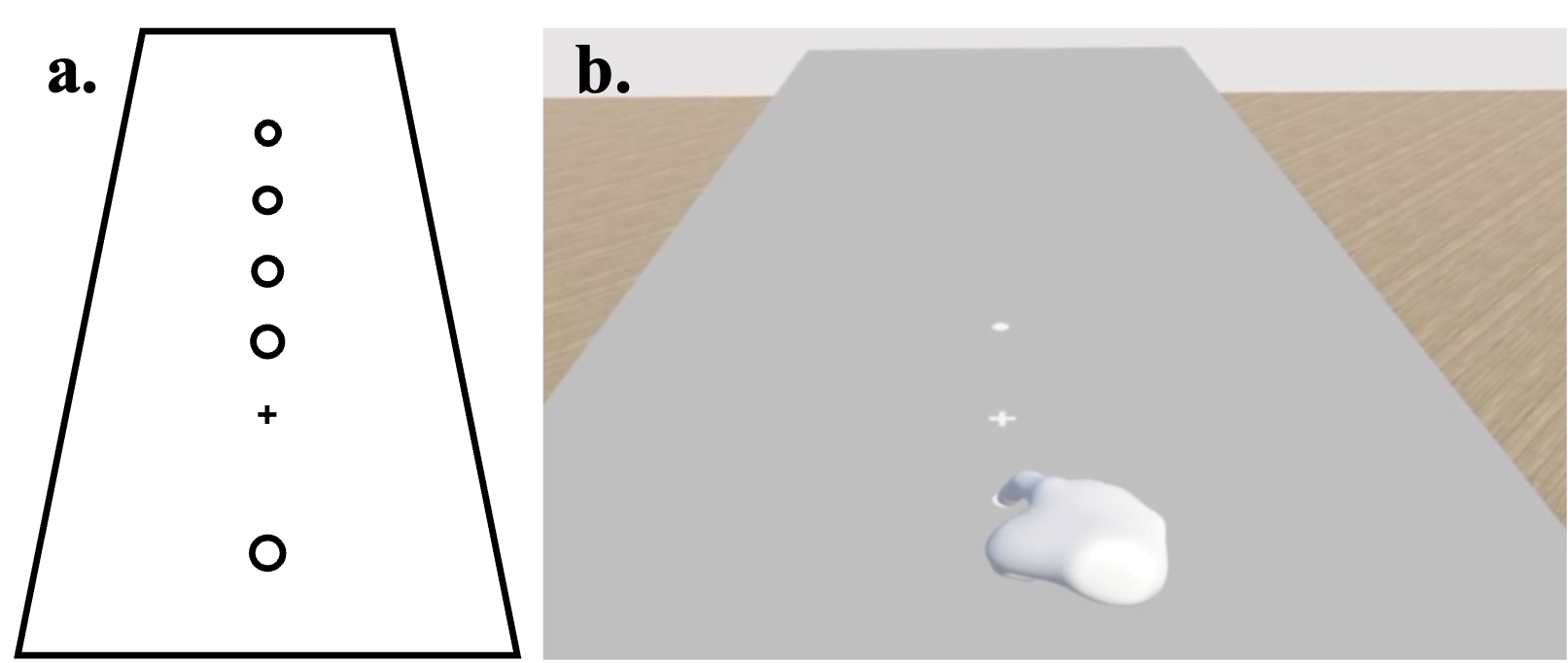}
  \caption{(a) The configuration of the home position, fixation cross, and potential target locations that were 20, 25, 30, and 35 cm away from the home position. (b) The setup of the virtual environment. Participants sat in front of a virtual table with a home position, a fixation cross, and a target. The real-time motion capture data from the IRED were used to animate the movement of a hand avatar. The height of the virtual table matched that of the physical table to provide accurate haptic feedback during the pointing movement.}
  \label{fig:procedures}
\end{figure}

The experiment was conducted in a controlled lab setting using an HTC VIVE Pro Eye HMD with a resolution of 1440 $\times$ 1600 pixels per eye and a refresh rate of 90 Hz. The HMD was powered by a custom desktop computer with an Intel Core i7 CPU, 32 GB RAM, and an NVIDIA RTX 3080 GPU. The virtual environment and interaction features were built in Unity and the experimental procedure was controlled using bmlTUX \cite{Bebko2020}. Figure~\ref{fig:procedures}a illustrates the schematic layout of the task, including a table, a central fixation cross, a circular home position, and four possible target locations (2 cm in diameter) positioned 20, 25, 30, and 35 cm from the home position along the depth axis, of which only one was presented on any given VR trial. The targets appeared directly in front of the participant and were aligned with the midline. In addition, baseline tests in the physical environment were conducted before and after the VR task using a physical foam board with the same home position and target locations. Procedures of the UR baseline tasks were controlled using PsychoPy \cite{Peirce2019}. Throughout the experiment, participants sat in front of a physical table with a same height as the virtual table to provide tactile feedback during the pointing movement.

A 3D virtual right hand, rendered in a neutral pointing posture, provided visual feedback of the participant's hand and finger positions (Figure \ref{fig:procedures}b). The hand was animated in real-time using data from a single infrared-emitting diode (IRED) placed on the right index fingertip. Kinematic data were collected at 250 Hz through an opto-electric motion capture system (3D Investigator, Optotrak, Waterloo, Canada) and streamed into Unity via a custom Python-Unity interface with a UDP socket. Using the method described in \cite{Warburton2023}, this system's native motion-to-photon latency was measured at 62 ms (SD = 8 ms). 

A custom calibration procedure was used to align the spatial reference frame of the motion capture system with that of the virtual environment. Using the HMD's video passthrough mode, participants viewed both the physical and virtual environments and sequentially pointed with their physical hand to three points on the virtual table. The positions of the participant's fingertip, recorded via Optotrak, and the corresponding virtual points were then used to compute a rigid transformation matrix consisting of translation and rotation components. This transformation matrix was subsequently used to animate the virtual hand by mapping kinematic data from physical space in the Optotrak reference frame to virtual space in the Unity reference frame.

To impose a visual delay on the virtual hand, transformed kinematic samples were buffered in a first-in/first-out queue with timestamps. Each entry contained the calibrated 3D hand position and the local clock time. At each frame, the current hand position was appended to the buffer. The renderer then selected the oldest entry that had remained in the buffer for at least the preset amount of time and used that position to update the virtual hand. For example, with a 200-ms delay the virtual hand position corresponded to the physical hand position 200 ms earlier, plus the additional 62 ms native delay.

\subsection{Procedures and Design}

After providing written informed consent, participants were seated at a physical table, and an IRED was affixed to the tip of their right index finger using medical tape. Participants first completed a series of baseline pointing trials in the physical environment (UR pre-test). There were four possible target locations positioned 20, 25, 30, and 35 cm from the home position along the depth axis. On each trial, participants placed their right index finger on the home position, after which the target number was announced and a beeping sound was played, signaling participants to point to the target as quickly and as accurately as possible. The four target locations constituted a block and were each presented once in randomized order. This block was repeated 14 times during the pre-test, resulting in 56 trials.

After completing the UR pre-test, participants were fitted with a VR HMD, with which they completed the spatial calibration procedure and then performed a series of pointing trials. As in the UR pre-test, participants began each trial with their right index finger placed on the home position. When ready, the experimenter initiated the trial, which first revealed the fixation cross. After 1,000 ms, a target appeared at one of the four possible locations, accompanied by a beeping signal that prompted movement initiation. Participants were instructed to point to the target as quickly and as accurately as possible. The experiment used a within-subject design blocked by visual delay (6 levels: 0, 10, 50, 100, 200, and 500 ms; note that these values refer to additional delay imposed beyond the native system latency, estimated at 62 ms). The order of the delay blocks was randomized across participants. Within each block, target distance (4 levels: 20, 25, 30, and 35 cm) was randomized across trials and repeated 14 times, resulting in 56 trials per block. In total, the VR task comprised 6 (visual delays) $\times$ 4 (target distances) $\times$ 14 (repetitions) = 336 trials.

Following the VR task, participants removed the HMD and performed the same pointing task in the physical environment as in the pre-test. That is, the UR post-test consisted of 14 repetitions of pointing to the four target locations, resulting in 56 trials.

\subsection{Data Reduction}

Kinematic data were processed using TAT-HUM, a Python-based kinematic analysis toolkit \cite{Wang2024TATHUM}. Missing data due to marker occlusion were imputed using linear interpolation if there were no more than 15 consecutive missing samples or, equivalently, over a 60 ms period. Trials with missing data segments that span more than 15 consecutive samples were discarded from the analysis (220 out of 6720 trials, or 3.27\% of total trials). 

Position data were smoothed using a low-pass Butterworth filter with a 250 Hz sampling frequency and a 10 Hz cutoff frequency to remove high frequency noise. Then, a 3-point finite difference method was used to derive movement velocity and acceleration, smoothed using the same Butterworth filter. Movement onset and termination were determined using a 50 mm/s velocity threshold along the primary movement direction in depth. Movement time (MT) is the amount of time between movement initiation and termination. Endpoint error (EE) is the signed deviations between the movement endpoint and the target along the primary movement direction. Signed EE preserved directional information regarding overshoot and undershoot, and the trial-level EE and MT were used in the subsequent statistical analyses to retain within-participant variability and maximize statistical power. Variable error (VE) is the standard deviation of EE across repetitions for each unique combination of conditions for each participant. 

Movement efficiency was quantified using throughput (TP) with the effective index of difficulty (${\mathrm{ID}}_e$; \cite{ISO2015, Batmaz2023}):

\begin{equation}
    {\mathrm{ID}}_e = \log_2\left(\frac{A_e}{W_e} + 1\right)
\end{equation}

\noindent where $A_e$ is the effective amplitude, defined as the mean reach distance from the home position along the depth axis, and $W_e$ is the effective width, $W_e = 4.133 \times \mathrm{VE}$. Because signed EE differs from endpoint position only by a constant target offset within each target distance condition, its standard deviation is unchanged. Consequently, the effective width term ($W_e$) equivalently reflects variability in reach distance or variability in signed EE (i.e., VE). TP, taken as an index of performance efficiency, was computed as

\begin{equation}
    \mathrm{TP} = \frac{\mathrm{ID}_e}{\mathrm{MT}}
\end{equation}

TP reflects the amount of spatial information transmitted per unit time, incorporating both movement precision through $W_e$ and movement duration through MT. Higher TP indicates greater performance efficiency, whereas lower TP can result from increased endpoint variability, longer movement time, or both. It is important to note that TP primarily summarizes speed and endpoint precision rather than signed endpoint bias. Although $A_e$ reflects the achieved movement amplitude, systematic directional error is not represented in TP in the same way as in EE. Thus, TP provides a useful aggregate index of speed-precision efficiency, but it cannot fully capture cases in which latency primarily increases signed endpoint error without a corresponding change in MT or VE.

\subsection{Statistical Analysis}

Statistical analyses were performed using a series of linear mixed-effects models (LMMs) in R (afex; \cite{Singmann2015}). For EE and MT, LMMs allowed the analysis of trial-level data while preserving within-participant variability, avoiding aggregation bias, and maximizing statistical power. These trial-level models included a participant-specific random intercept and an uncorrelated random slope for target distance, allowing participants to vary in both baseline performance and distance scaling while simplifying the random-effects structure. In contrast, VE and TP are aggregated measures defined across repetitions within each participant and condition and therefore could not be modeled at the trial level. Accordingly, analyses of VE and TP were conducted on condition-level summaries for each participant. Because this aggregation substantially reduced the number of observations, models including participant-specific random slopes for target distance yielded singular fits. Therefore, the random-effects structure for VE and TP was simplified to include participant-specific random intercepts only. Model comparison favored this simpler structure.

The primary dependent variables were EE, MT, VE, and TP as measures of movement accuracy, timing, precision, and efficiency, respectively. Two sets of models were fitted to align the analyses with the research questions. First, modality models compared performance across the UR pre-test, the VR-0 ms condition, and the UR post-test to assess baseline differences between immersive VR and the physical environment (\textbf{H1}). This comparison captured the combined influence of the HMD's inherent hardware characteristics, the data pipeline's native latency, and other VR-specific factors. Second, visual delay models were restricted to VR trials and examined how additional visual delay affected performance across the 0, 10, 50, 100, 200, and 500 ms delay conditions (\textbf{H2}), and whether spatial and temporal outcomes showed dissociable response patterns (\textbf{H3}). For both sets of analyses, target distance $D$ was modeled as a centered, rescaled numeric variable $D_c$ for interpretability and numerical stability:

\begin{equation}
    D_c = \frac{D_i - \bar{D}}{5}
    \label{eq:dist_scaling}
\end{equation}

The modality models coded modality as a categorical fixed effect with three levels: UR pre-test (reference level), VR-0 ms, and UR post-test. Fixed effects included modality, target distance ($D_c$), and their interaction. The visual delay models were restricted to VR trials and coded visual delay as a categorical within-subject factor with six levels (0, 10, 50, 100, 200, and 500 ms; reference level: 0 ms). Delay was treated categorically because the levels were unevenly spaced and the relationships between delay and the dependent measures were expected to be nonlinear and potentially non-monotonic. Fixed effects included visual delay, target distance ($D_c$), and their interaction.

All models were estimated with restricted maximum likelihood (REML) using the BOBYQA (bound optimization by quadratic approximation) optimizer. Omnibus tests for fixed effects were conducted using Type III analysis of variance (ANOVA) with Satterthwaite degrees of freedom. Effect sizes were reported as partial $\eta^2$ ($\eta_p^2$), computed for fixed effects using the \texttt{effectsize} package \cite{BenShachar2020}. Estimated marginal means (EMMs) with Tukey-adjusted pairwise comparisons were computed. For the modality model, contrasts focused on UR pre-test vs.\ VR-0 ms, VR-0 ms vs.\ UR post-test, and UR pre-test vs.\ UR post-test. For the visual delay model, pairwise contrasts focused primarily on comparisons against the 0-ms condition, and EMM trends were used to estimate the slope of target distance across delay conditions.

\section{Results and Discussion}

% \begin{figure*}[t]
%   \centering
%   \includegraphics[width=\textwidth]{Figures/manuscript_delay_all_hci.png}
%   \caption{Effects of visual delay on pointing performance. (\textbf{A}) Endpoint error (cm), (\textbf{B}) movement time (s), (\textbf{C}) variable error (cm), and (\textbf{D}) throughput (bits/s). Within each panel, the left subpanel shows the modality comparison between UR Pre-test, VR-0 ms, and UR Post-test, and the right subpanel shows the effect of additional visual delay in VR (0--500 ms). The 0 ms condition is repeated intentionally because it serves as the bridge between the modality comparison and the within-VR delay comparison. Error bars are bootstrapped 95\% confidence intervals.}
%   \label{fig:results}
% \end{figure*}

\subsection{VR Baseline Differences from the Physical Environment}

\begin{figure}[tb]
  \centering
  \includegraphics[width=\columnwidth]{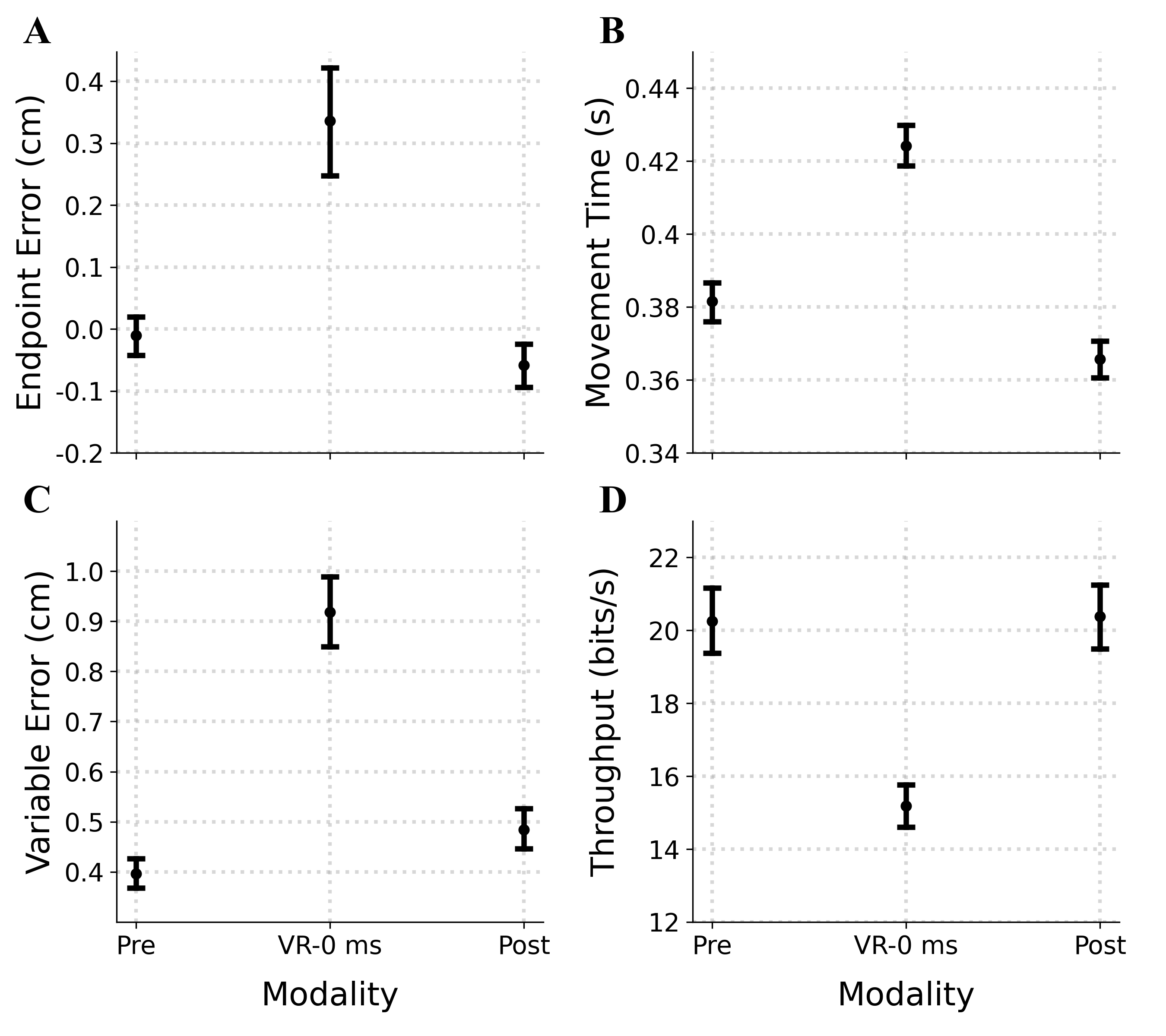}
  \caption{Baseline differences in pointing performance between the physical environment and VR. (\textbf{A}) Endpoint error (cm), (\textbf{B}) movement time (s), (\textbf{C}) variable error (cm), and (\textbf{D}) throughput (bits/s) are shown for the UR pre-test, the VR-0 ms condition, and the UR post-test. Error bars are bootstrapped 95\% confidence intervals.}
  \label{fig:modality}
\end{figure}

To isolate baseline differences between immersive VR and the physical environment, the modality models were used to compare pointing performance across the UR pre-test, the VR-0 ms condition, and the UR post-test (Figure~\ref{fig:modality}).

\subsubsection{Endpoint Error}

EE was higher in VR than in the physical environment (Figure~\ref{fig:modality}A). Omnibus tests showed significant main effects of target distance, $F(1, 18.5) = 15.24, p < .001, \eta_{p}^2 = .45$, and modality, $F(2, 2927.1) = 59.56, p < .001, \eta_{p}^2 = .04$, as well as a significant interaction, $F(2, 2925.9) = 66.03, p < .001, \eta_{p}^2 = .04$. EE in the VR-0 ms condition was significantly greater than that in the UR pre-test (mean difference = +0.309 cm, $p < .001$) and the UR post-test (mean difference = +0.38 cm, $p < .001$). Although EE shifted slightly toward undershooting from the UR pre-test to the UR post-test, the difference between them failed to reach significance (mean difference = -0.072 cm, $p = 0.054$). Together, these estimates indicate that VR produced a robust increase in EE even without additional delay, reflecting noticeable overshooting at the mean target distance.

\begin{figure}[tb]
  \centering
  \includegraphics[width=\columnwidth]{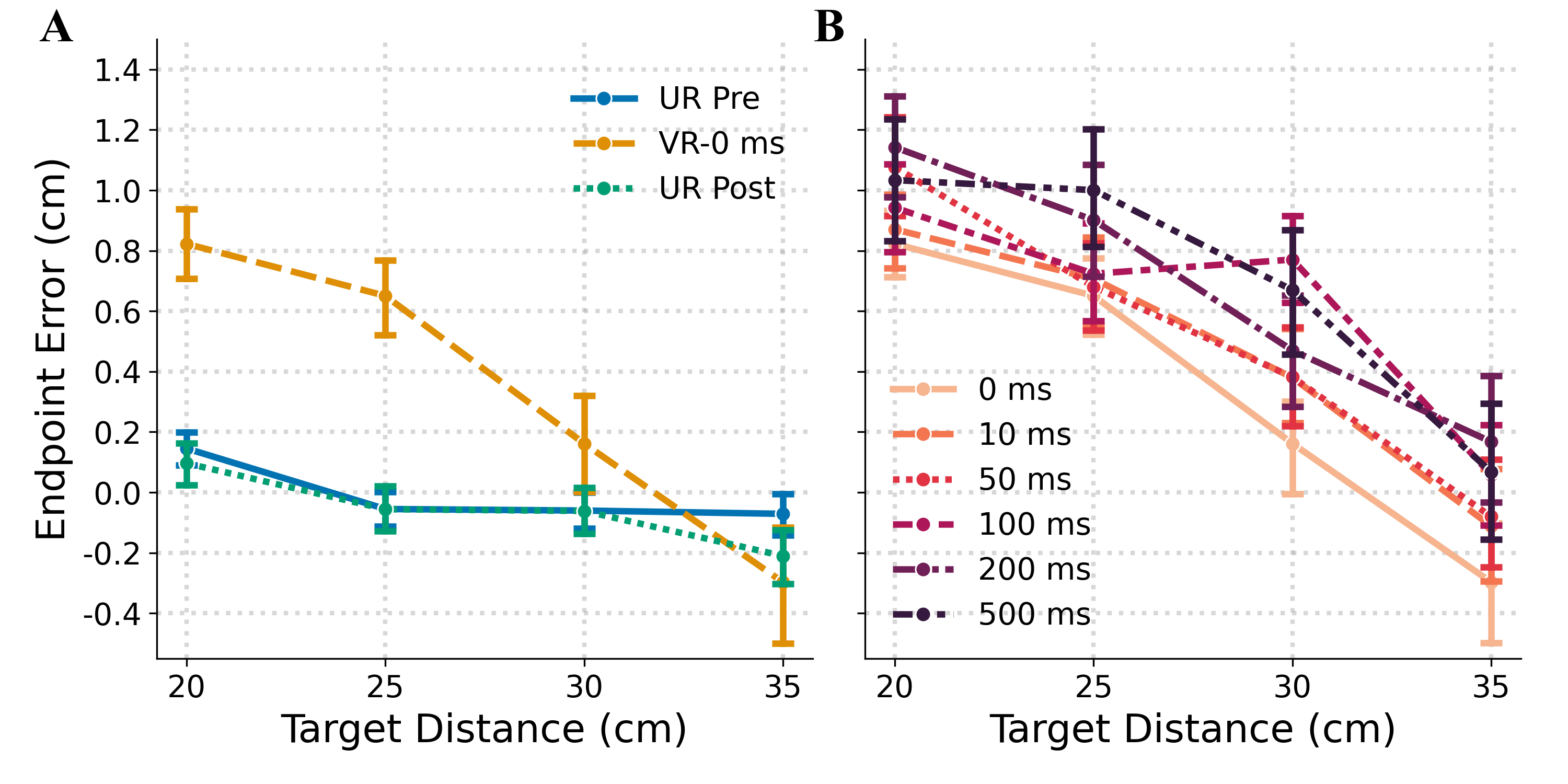}
  \caption{Distance-dependent scaling of endpoint error across (\textbf{A}) modality conditions and (\textbf{B}) added visual delay conditions. Endpoint error is shown as a function of target distance (20--35 cm). In the modality comparison, endpoint error remained relatively stable across target distance in the physical environment, whereas the VR-0 ms condition exhibited a steeper negative slope with increasing distance, consistent with depth compression related to the vergence-accommodation conflict (VAC). In the delay comparison, added visual delay shifted endpoint error upward but did not substantially alter the distance-dependent trend. Error bars represent bootstrapped 95\% confidence intervals.}
\label{fig:vac_dist}
\end{figure}

To unpack the significant modality-by-distance interaction, distance trends were compared across modality conditions (Figure~\ref{fig:vac_dist}A). The distance trend in the UR pre-test was small and not significant (slope = -0.039 per 5 cm, $p = .40$). The distance slope in VR was more negative than in the pre-test ($\Delta = -0.343$, $p < .001$), yielding an overall negative trend of -0.382 per 5 cm. The UR post-test did not differ significantly from the pre-test in slope ($\Delta = -0.029$, $p = .38$). The steeper negative slope in VR indicates that the direction and magnitude of EE changed more strongly with target distance in VR than in the physical environment. This observation is consistent with VAC-related depth compression observed in previous studies \cite{Wang2026Displays,Wang2024bVR}: as target distance increases, perceived depth is increasingly compressed, producing progressively larger undershoots.

\subsubsection{Movement Time}

MT was longer in VR than in the physical environment (Figure~\ref{fig:modality}B). The modality model showed significant main effects of target distance, $F(1, 20.0) = 295.76$, $p < .001$, $\eta_{p}^2 = .94$, and modality, $F(2, 2916.0) = 484.17$, $p < .001$, $\eta_{p}^2 = .25$, as well as a significant interaction, $F(2, 2904.8) = 9.81$, $p < .001$, $\eta_{p}^2 = .007$. MT was significantly longer in the VR-0 ms condition than in the UR pre-test (mean difference = +49 ms, $p < .001$) and the UR post-test (mean difference = +62 ms, $p < .001$). MT in the UR post-test was significantly shorter than in the pre-test (mean difference = -13 ms, $p < .001$). The increased MT in VR relative to the physical baselines was consistent with previous studies showing longer MT in VR than in the physical environment \cite{Wang2025BBR}. The shorter MT in the UR post-test also indicates a small practice-related decrease from pre- to post-test.

Post hoc tests for the modality-by-distance interaction showed that MT increased with target distance in all modalities (all $p < .001$). However, this distance-related increase was significantly steeper in VR than in the pre-test ($\Delta b = 0.005$, $p = .009$) and the post-test ($\Delta b = 0.008$, $p < .001$), whereas the pre- and post-test slopes did not differ ($\Delta b = 0.003$, $p = .343$). Thus, although participants took longer to reach farther targets in all conditions, the effect of distance on MT was amplified in VR. This pattern suggests that reaching to more distant targets imposed a greater temporal cost in VR than in the physical environment.

\subsubsection{Variable Error}

Endpoint variability was also elevated in VR (Figure~\ref{fig:modality}C). The fitted modality model revealed significant main effects of target distance, $F(1, 199.8) = 9.82$, $p < .01$, $\eta_{p}^2 = .05$, and modality, $F(2, 203.2)=156.33$, $p < .001$, $\eta_{p}^2 = .61$. The interaction was not significant, $F(2, 199.8) = 0.15$, $p = .86$, $\eta_{p}^2 = .001$. VE in the VR-0 ms condition was significantly greater than in the UR pre-test (mean difference = +0.52 cm, $p < .001$) and the UR post-test (mean difference = +0.43 cm, $p < .001$). VE in the UR post-test was also significantly greater than in the UR pre-test (mean difference = +0.087 cm, $p < .05$). Combined with the longer MT in VR, the increased VE suggests that the elevated EE in VR was not attributable to a simple speed-accuracy trade-off.

\subsubsection{Throughput}

Movement efficiency was lower in VR compared to the UR baselines (Figure~\ref{fig:modality}D). The modality model showed significant main effects of target distance, $F(1, 199.2) = 86.02$, $p < .001$, $\eta_{p}^2 = .30$, and modality, $F(2, 200.2) = 230.72$, $p < .001$, $\eta_{p}^2 = .70$. Their interaction was not significant, $F(2, 199.2) = .99$, $p = .37$, $\eta_{p}^2 = .010$. Planned contrasts showed that TP in the VR-0 ms condition was significantly lower than in the UR pre-test (mean difference = 5.13 bits/s, $p < .001$) and the UR post-test (mean difference = 5.26 bits/s, $p < .001$). The difference between the UR pre- and post-tests was not significant (mean difference = 0.13 bits/s, $p = .89$). This reduction in TP in VR-0ms mirrors the elevated VE and MT reported earlier.

\subsection{Latency-Dependent Dissociation Between Accuracy and Timing}

\begin{figure}[tb]
  \centering
  \includegraphics[width=\columnwidth]{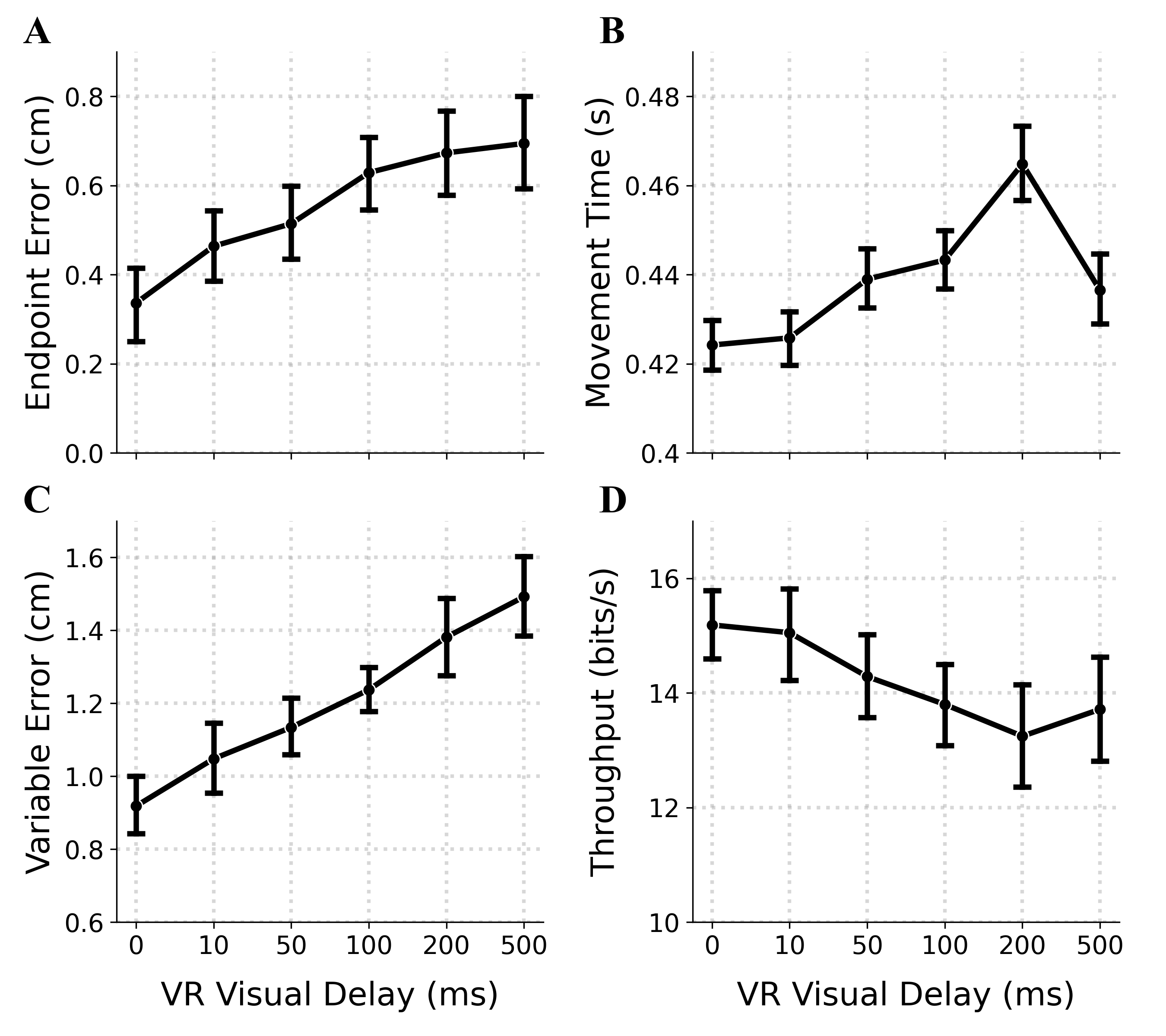}
  \caption{Effects of added visual delay on pointing performance in VR. (\textbf{A}) Endpoint error (cm), (\textbf{B}) movement time (s), (\textbf{C}) variable error (cm), and (\textbf{D}) throughput (bits/s) are shown across additional visual delay conditions from 0 to 500 ms. Error bars are bootstrapped 95\% confidence intervals.}
  \label{fig:delay}
\end{figure}

The preceding analyses showed that, even without added interaction latency, performance in VR was degraded relative to UR: movements were less accurate, longer, more variable, and less efficient. These baseline differences likely reflect the combined influence of several VR-specific factors, including VAC, native motion-to-photon latency, and HMD ergonomics. To isolate the effect of interaction latency, additional delays were introduced between the participant's physical hand movement and the displayed virtual hand. Movement accuracy, timing, consistency, and efficiency were then compared across delay conditions (Figure~\ref{fig:delay}).

\subsubsection{Endpoint Error}

There was a front-loaded, nonlinear relationship between visual delay and EE (Figure~\ref{fig:delay}A). The visual delay model showed significant main effects of visual delay, $F(5, 6322.3) = 11.08$, $p < .001$, $\eta_{p}^2 = .009$, and target distance, $F(1, 19.0) = 27.74$, $p < .001$, $\eta_{p}^2 = 0.59$. Their interaction was not significant, $F(5, 6322.3) = 1.47$, $p = .20$, $\eta_{p}^2 = .001$. At the mean target distance, EE at 0 ms delay was 0.33 cm (SE = 0.11). Relative to 0 ms, EE increased across delay levels by +0.13 cm at 10 ms ($p < .05$), +0.18 cm at 50 ms ($p < .01$), +0.29 cm at 100 ms ($p < .001$), +0.33 cm at 200 ms ($p < .001$), and +0.36 cm at 500 ms ($p < .001$). Initially, introducing 10 ms additional delay resulted in a considerable increase in EE by +0.13 cm (\~39\% relative to the 0-ms level), whereas increasing delay from 100 ms to 200 ms added only 0.04 cm to EE (\~6\% relative to the 100-ms level). This pattern suggests a decelerating trend, in which smaller delays produced relatively large increases in EE, whereas additional increases at longer delays produced progressively smaller changes.

The main effect of target distance indicated that EE decreased with distance (slope = $-0.39$ per 5 cm, $p < .001$; Figure~\ref{fig:vac_dist}B). This negative trend replicated the distance-dependent undershooting expected from VAC-related depth compression \cite{Wang2026Displays,Wang2024bVR}. Added visual delay shifted EE upward across target distances but minimally altered this distance-dependent trend: only the 100 ms condition showed a reliably less negative slope ($\Delta = 0.13$ per 5 cm, $p < .05$), whereas the 10, 50, 200, and 500 ms conditions did not differ from the 0 ms baseline (all $p > .23$). Together, these estimates indicate that visual delay had a largely additive effect on EE, with little evidence that delay systematically modulated the VAC-related distance-error relationship.

\subsubsection{Movement Time}

Unlike EE that increased even at 10 ms, MT remained unchanged at short delays and increased only at longer delays (Figure~\ref{fig:delay}B). The omnibus test revealed significant main effects of visual delay, $F(5, 6322.1) = 49.49$, $p < .001$, $\eta_{p}^2 = 0.04$, and target distance, $F(1, 18.9) = 184.97$, $p < .001$, $\eta_{p}^2 = .91$. The interaction was not significant, $F(5, 6322.3) = 1.23$, $p = .29$, $\eta_{p}^2 = 0.001$. Planned contrasts against the 0 ms condition showed that MT increased by +14 ms at 50 ms ($p < .001$), +17 ms at 100 ms ($p < .001$), +39 ms at 200 ms ($p < .001$), and +11 ms at 500 ms ($p < .001$). The 10 ms delay had essentially no effect on MT ($p \approx 1.00$), which was in stark contrast EE that showed the largest increase at 10 ms. This finding indicates that the spatial and temporal consequences of delay emerged differently. 

Further pairwise comparisons clarified the non-monotonic structure of this effect: MT did not differ between 50 and 100 ms ($p = .79$), but increased from 50 to 200 ms ($p < .001$) and from 100 to 200 ms ($p < .001$). Across the 50--200 ms range, this corresponds to an approximate increase of 2 ms in MT for every 10 ms of added visual delay. This pattern suggests that participants may have slowed their physical hand movement to partially compensate for delayed visual feedback. However, MT at 500 ms was lower than at 200 ms ($p < .001$) and did not differ from the 50 or 100 ms conditions (both $p > .32$), suggesting a breakdown of compensatory slowing at the longest delay. Because the imposed lag at 500 ms exceeded the average MT (Figure~\ref{fig:delay}B), participants may have reduced reliance on delayed visual feedback and relied more heavily on feedforward control.

\subsubsection{Variable Error}

The effect of added visual delay on VE broadly mirrored its effect on EE (Figure~\ref{fig:delay}C). The visual delay model showed significant main effects of target distance, $F(1, 449) = 14.86$, $p < .001$, $\eta_{p}^2 = .03$, and visual delay, $F(5, 449) = 32.42$, $p < .001$, $\eta_{p}^2 = .27$. Their interaction was not significant, $F(5, 449) = 0.78$, $p = .56$, $\eta_{p}^2 = .009$. As with EE, VE increased as a function of additional delay relative to 0 ms: +0.13 cm at 10 ms ($p = .07$), +0.22 cm at 50 ms ($p < .001$), +0.32 cm at 100 ms ($p < .001$), +0.46 cm at 200 ms ($p < .001$), and +0.57 cm at 500 ms ($p < .001$).

Together with the MT results, this pattern suggests that the delay-related MT increases were not attributable to a simple speed-accuracy trade-off. Specifically, VE increased with delay even though longer MT would ordinarily be expected to reduce endpoint variability under a speed-accuracy trade-off account. VE rose monotonically from 0 to 500 ms without a reliable interaction with target distance, indicating that delay increased endpoint variability largely independent of reach amplitude. Taken together with the EE analysis, these results indicate that delay inflated both the bias and variance of movement endpoints. The absence of a corresponding reduction in VE when MT increased from 50 to 200 ms, as well as the continued increase in VE at 500 ms even as movement appeared to shift toward more feedforward control, is more consistent with partial compensation under delayed feedback. Specifically, participants may have slowed their movements once the delay became perceptually detectable, but this slowing would not fully eliminate the spatial mismatch between the physical hand and the delayed virtual hand. As a result, larger delays could continue to increase both endpoint bias and variability even when movement time increased. Given that delay increased endpoint variability while also prolonged MT, TP was analyzed next to determine how these changes jointly affected overall movement efficiency.

\subsubsection{Throughput}

TP decreased with added visual delay, indicating reduced overall movement efficiency (Figure~\ref{fig:delay}D). The visual delay model showed significant main effects of target distance, $F(1, 449) = 119.46$, $p < .001$, $\eta_{p}^2 = .21$, and visual delay, $F(5, 449) = 21.03$, $p < .001$, $\eta_{p}^2 = .19$. Their interaction was not significant, $F(5, 449) = 0.67$, $p = .65$, $\eta_{p}^2 = .007$. Compared to the 0 ms condition, TP did not differ at 10 ms ($p = .99$), but decreased by 0.90 bits/s at 50 ms ($p < .001$), 1.39 bits/s at 100 ms ($p < .001$), 1.94 bits/s at 200 ms ($p < .001$), and 1.47 bits/s at 500 ms ($p < .001$). However, this decrease did not continue uniformly at longer delays. While TP continued to decline from 50 to 200 ms ($p < .001$), additional pairwise comparisons among the higher delay levels showed no significant differences between the 100, 200, and 500 ms conditions (all $p > .18$), indicating that efficiency loss largely plateaued at higher delays. This pattern suggests that the primary loss in efficiency occurred between baseline and moderate delays, with limited additional degradation at longer delays.

\section{General Discussion}

The current study examined how interaction latency shapes visually guided manual pointing in immersive VR by separating baseline VR effects from experimentally imposed delays. Consistent with \textbf{H1}, performance in VR was degraded relative to UR even in the absence of additional delay: movements were less accurate, longer, more variable, and less efficient. This baseline deficit likely reflects the combined influence of VR-specific factors, including VAC-related depth compression \cite{Wang2026Displays, Wang2024bVR}, native system latency \cite{Warburton2023}, and altered sensorimotor coupling \cite{ManzoneUnderReview, Wang2025BBR}. Consistent with \textbf{H2}, increasing interaction latency within VR further impaired performance. Added delay systematically increased EE and VE, reduced TP, and altered MT at longer delays. Importantly, the findings also supported \textbf{H3} because the performance measures exhibited dissociable response patterns as interaction latency increased. Spatial measures (EE and VE) and temporal/efficiency measures (MT and TP) responded differently to increasing delay. Both EE and VE increased with added delay, but EE showed evidence of earlier disruption, increasing significantly even at the shortest delay condition (10 ms). Although both measures increased across delay levels, the increase in EE appeared to diminish at longer delays, whereas VE showed a more gradual increase across the delay range. In contrast, MT and TP were initially unchanged and only changed reliably at longer delays. This dissociation indicates that commonly used performance measures are not interchangeable when evaluating latency-sensitive VR interactions, because different measures revealed latency-related disruption at different delay magnitudes.

The dissociation among spatial, temporal, and efficiency measures suggests that interaction latency can perturb online movement regulation before producing overt changes in movement timing or overall efficiency. Earlier work in motor control showed that delayed visual feedback primarily disrupts online visual guidance during movement execution rather than the initial motor command itself \cite{Smith1980}. In that study, a 66 ms visual feedback delay produced the largest spatial errors at intermediate movement times (approximately 350--450 ms), whereas very rapid movements (approximately 150 ms) showed smaller effects because there was insufficient time for delayed visual information to substantially influence online control (see also \cite{Elliott2001}). In the current study, changes in MT emerged only at longer delays, suggesting that participants may have adopted a more conservative control strategy when the temporal mismatch between movement and visual feedback became sufficiently large. Thus, spatial error may provide an earlier index of disrupted online control, whereas MT may reflect a later compensatory response to that disruption.

Findings from the current study align with prior work showing that the behavioral consequences of interaction latency depend on both delay magnitude and movement demands. In desktop pointing and steering tasks, Friston et al. \cite{Friston2015} reported that MT increased only once latency exceeded a functionally meaningful range, with little effect at shorter delays. Similar threshold-like effects have been observed in VR, such that latencies greater than approximately 75 ms degraded motor performance and embodiment-related judgments, with larger delays producing progressively greater deterioration \cite{Waltemate2016}. Hoyet et al. \cite{Hoyet2019} further demonstrated that latency costs depend on movement speed: little degradation was observed up to approximately 120 ms at slow tracking speeds (350 mm/s), whereas performance declined around 80 ms at medium and high (500--650 mm/s) speeds. The MT findings from the current study are consistent with this literature. In this context, the emergence of MT costs between 50 and 100 ms is consistent with a speed-dependent threshold account, wherein MT remained largely unchanged at the shortest delays and increased primarily at moderate delays.

Together, these findings suggest that low or sub-threshold interaction latency should not be assumed behaviorally negligible simply because overt timing changes are absent. EE increased even at the shortest delays despite little or no corresponding change in MT, indicating that visually guided action can be disrupted before users visibly alter their movement strategy. This dissociation is particularly important for HCI because latency effects are often summarized using movement time or Fitts-style efficiency measures (e.g., \cite{Friston2015,Hoffmann1992,Yamanaka2024}), although recent work has begun to model latency-related changes in endpoint distributions to better explain selection behavior in VR \cite{Wei2026}. In the current study, TP exhibited a similar pattern to MT, showing little change at the shortest delay and then decreased at moderate delays. However, this efficiency loss largely plateaued at higher delays. Thus, although TP captured an overall decline in performance efficiency by combining endpoint variability and movement time, its plateau at higher delays obscured the qualitatively different response profiles of EE, VE, and MT. 

One interpretation of the dissociation between spatial and temporal performance measures is that participants adopted a partial compensation strategy once interaction latency became behaviorally meaningful. Specifically, participants may have increased MT to allow more time for feedback-based control. However, the additional time was insufficient to eliminate the temporal mismatch between delayed visual feedback and the current physical hand position. Larger delays therefore continued to elevate both endpoint bias and endpoint variability. Together, these findings suggest that delayed visual feedback alters online control strategies without fully restoring veridical visually guided performance.

The current study has implications for the design and evaluation of VR systems in terms of motion-to-photon latency. Because some degree of latency is unavoidable in immersive systems, prior work has largely focused on minimizing end-to-end delay through improved tracking, rendering, and prediction techniques \cite{Hou2020,Hou2019,LaValle2014}. This stragegy remains essential for usability and comfort, particularly because view-contingent latency contributes to visually induced motion sickness and perceptual instability \cite{Keshavarz2015,Lackner2000,Stauffert2018,Stauffert2020}. However, the current results suggest that the design problem for manual interaction extends beyond simply reducing delay magnitude. Even very small additional delays increased EE despite producing little observable change in MT. This pattern indicates that users may continue to rely on delayed visual feedback even when latency is not sufficiently large to trigger overt compensatory behavior. Accordingly, latency evaluation in VR should consider not only whether users can perceive delay, but also whether sub-threshold delay still alters action performance.

More broadly, the dissociation between spatial, temporal, and efficiency measures highlights an important methodological consideration for HCI research on immersive 3D virtual environments. Although VR introduces additional spatial and sensorimotor complexity \cite{Wang2023VisualCognition,Wang2024JOV}, many design and evaluation approaches remain rooted in traditional 2D interaction paradigms (e.g., \cite{Zhang2026}), emphasizing movement time, throughput, and selection success rate (e.g., \cite{Wei2026,Yamanaka2024}). These measures are useful for characterizing overall interaction efficiency, but they may not fully capture how immersive VR affects movement control when interaction depends on continuous spatial guidance, embodied action, and online visuomotor regulation. This issue is especially relevant for applied VR contexts, such as surgical or engineering training (e.g., \cite{Giraldo2025,Wang2025ASEE}), where the goal is not only to select targets efficiently but also to support functional movements that approximate real-world 3D interactions. Although immersive VR cannot completely replicate perceptuomotor experience in the physical environment \cite{ManzoneUnderReview, Wang2025BBR}, it remains a valuable tool for training interactions that may otherwise be difficult to access because of cost, safety, or logistical constraints. Therefore, evaluating latency-sensitive VR interactions requires measures that capture both movement timing and spatial accuracy, particularly when the task involves actions that are more representative of real-world 3D performance.

\subsection{Limitations and Future Directions}

One limitation of the current study is that it did not directly measure participants' behavioral sensitivity to interaction latency or their subjective awareness of delayed visual feedback. Consequently, the interpretation that participants may have adopted a partial compensatory strategy once latency became behaviorally meaningful remains inferential rather than empirically verified. In particular, the current findings cannot determine when participants became consciously aware of the mismatch between the physical and virtual hand, whether perceived latency aligned with the delay levels at which different performance measures began to change, or how such awareness influenced the control strategies used to maintain fast and accurate visually guided movements. Future work should therefore combine performance-based measures with psychophysical estimates of delay detectability and judgments of perceived action-feedback mismatch. Such approaches would help clarify how perceptual sensitivity to interaction latency relates to spatial error, movement timing, and compensatory control during visually guided action in VR. They would also allow future studies to examine whether individual differences in latency sensitivity predict when spatial disruption emerges and when compensatory strategies are adopted.

A second limitation is that the current task involved a relatively constrained visually guided pointing movement under controlled laboratory conditions. Further, the effector of the participant in the virtual environment was a single disembodied hand that did not dynamically change in orientation as the hand moved through space. Although this simplified paradigm enabled precise manipulation of interaction latency and isolation of spatial and temporal performance measures, real-world VR interactions often involve more complex full-body movements, object manipulation, continuous feedback loops, and multisensory integration. Future work should therefore investigate whether the dissociable latency effects observed in the current study generalize to more naturalistic interaction scenarios and applied VR contexts.

More broadly, the current findings raise questions not only about the immediate behavioral effects of latency, but also about how users adapt to altered sensorimotor mappings in immersive virtual environments. These observations also raise the possibility that latency in VR may not always need to be treated solely as a constraint to minimize. From a seamful design perspective \cite{Chalmers2003a,Chalmers2003b}, making interaction latency perceptually explicit during brief onboarding or acclimation periods may facilitate adaptation to the altered temporal mapping between action and visual feedback in immersive environments. Such approaches would need to be implemented carefully to avoid discomfort, particularly because view-contingent latency is strongly associated with visually induced motion sickness. Future work should investigate whether controlled exposure to effector-specific latency can facilitate adaptation to VR-specific sensorimotor mappings without compromising long-term usability or comfort.

\section{Conclusion}

The current study examined how interaction latency alters visually guided manual pointing in immersive VR by separating baseline VR effects from experimentally imposed delays. Relative to unmediated performance, VR movements were less accurate, longer, more variable, and less efficient even without additional delay. Increasing interaction latency further impaired performance, but these effects were not uniform across measures. Spatial measures, particularly endpoint error and endpoint variability, showed earlier and more consistent degradation than timing and efficiency measures, indicating that delayed visual feedback disrupts online movement regulation before overt compensatory changes in movement timing emerge. These findings demonstrate that latency-sensitive VR interactions cannot be fully characterized using movement time or throughput alone. Although aggregate efficiency measures captured overall reductions in performance, they obscured important dissociations between the spatial and temporal consequences of delayed visual feedback. Evaluating both movement timing and spatial accuracy may therefore provide a more complete understanding of how interaction latency alters visually guided action in immersive environments. More broadly, the present findings highlight the importance of considering the unique spatial and sensorimotor demands of immersive 3D interaction when designing and evaluating VR systems. By clarifying how interaction latency shapes manual control in immersive VR, this study provides practical guidance for the evaluation of latency-sensitive VR interactions and establishes a foundation for future work on adaptation, online control, and performance assessment in immersive environments.

\section*{Acknowledgment}

The authors would like to thank Gavin Gibbs for the assistance in data collection. This project draws upon research supported by the Government of Canada's New Frontiers in Research Fund (NFRF) and Natural Sciences and Engineering Research Council (NSERC).

During the preparation of this work, the authors used OpenAI ChatGPT to assist with language editing, manuscript organization, and feedback on presentation. The authors reviewed and edited all generated content and take full responsibility for the final manuscript.

\end{document}